%% file: 00-main.tex
\documentclass[sigplan,10pt,nonacm]{acmart}

\renewcommand\footnotetextcopyrightpermission[1]{}

\input{imports}

\begin{document}
\date{}

\title{FastEEP: Fast Intra-Instance Elastic Scaling for Expert-Parallel MoE Serving}

\author{
Yunfei Feng\footnotemark[1],
Jingkai He\footnotemark[1],
Shiqi Liu,
Cheng Zhang,
Dong Du,
Erhu Feng\footnotemark[2]
}

\affiliation{
\institution{
Institute of Parallel and Distributed Systems (IPADS), Shanghai Jiao Tong University
}
\country{}
}

\settopmatter{printacmref=false}
\settopmatter{printfolios=true}
\setcopyright{none}

\input{01-abstract}
\maketitle
\begingroup
\renewcommand{\thefootnote}{\fnsymbol{footnote}}
\footnotetext[1]{Yunfei Feng and Jingkai He contributed equally to this work and are co-first authors.}
\footnotetext[2]{Correspond to: \href{mailto:fengerhu1@sjtu.edu.cn}{fengerhu1@sjtu.edu.cn}}
\endgroup

\pagestyle{plain}

\input{02-introduction}

\input{03-background}

\input{04-overview}
\input{05-design}
\input{06-evaluation}
\input{07-discussion}
\input{08-related}
\input{09-conclusion}

\afterpage{\balance}
\bibliographystyle{ACM-Reference-Format}
\bibliography{refs}

\end{document}

%% file: imports.tex
\usepackage{booktabs}
\usepackage{multirow}
\usepackage{array}
\usepackage{subcaption}
\usepackage{xspace}
\usepackage{enumitem}
\usepackage{afterpage}

\DeclareCaptionFont{myfont}{\fontsize{9pt}{10.8pt}\selectfont}
\newcommand{\tabsetup}{%
  \footnotesize
  \setlength{\tabcolsep}{4pt}%
  \renewcommand{\arraystretch}{1.18}}
\newcolumntype{L}[1]{>{\raggedright\arraybackslash}p{#1}}
\newcolumntype{C}[1]{>{\centering\arraybackslash}p{#1}}
\newcolumntype{R}[1]{>{\raggedleft\arraybackslash}p{#1}}
\newcommand{\colhd}[1]{\textbf{#1}}

\newcommand{\naa}{\textmd{--}}

\newif\ifsvgpkg
\IfFileExists{svg.sty}{%
  \immediate\write18{rm -f inkscape-ok.tex;
    inkscape --version >/dev/null 2>&1 && echo ok > inkscape-ok.tex}%
  \IfFileExists{inkscape-ok.tex}{\usepackage{svg}\svgpkgtrue}{}%
}{}

\newcommand{\figmissing}[1]{%
  \fbox{\parbox[c][0.28\linewidth][c]{0.95\linewidth}{\centering\footnotesize
    \texttt{imgs/#1.svg}\\[2pt]
    export a PDF from draw.io, or build with the \texttt{svg} package}}}

\newcommand{\incfig}[1]{%
  \IfFileExists{imgs/#1.pdf}%
    {\includegraphics[width=\linewidth]{imgs/#1.pdf}}%
    {\ifsvgpkg\includesvg[width=\linewidth]{imgs/#1.svg}\else\figmissing{#1}\fi}}

\newcommand{\sys}{FastEEP\xspace}

%% file: 01-abstract.tex

\begin{abstract}
Large-scale expert parallelism (EP) has become the standard deployment
configuration for Mixture-of-Experts (MoE) models in production.
Meanwhile, the highly dynamic nature of LLM workloads requires providers to continuously adjust serving capacity to match fluctuating demand. 
Scaling MoE serving cannot be efficiently achieved by simply adding or removing serving instances; instead, it often requires resizing the EP group within an existing instance, which we refer to as intra-instance scaling.
Existing inference engines primarily rely on instance-level scaling, which destroys and reconstructs entire serving instances, incurring substantial scaling overhead.

We present \sys (Fast Elastic Expert Parallel), a fast intra-instance scaling system for MoE serving
based on the principle of \textbf{Partial Reuse}. \sys reuses existing
runtime state across worker execution, GPU memory, and communication by
separating reusable execution structures from the resources activated by
the current EP configuration. This separation enables coordinated resource
reclamation and reconfiguration while preserving costly-to-reconstruct state,
allowing fast EP scaling without full instance teardown and reconstruction.
Our evaluation shows that \sys achieves $4.5\times$--$11.7\times$ faster
scaling than the existing vLLM-based approach across different models,
enabling efficient, fine-grained scaling of MoE serving.

\end{abstract}

%% file: 02-introduction.tex

\section{Introduction}
\label{sec:intro}

Mixture-of-Experts (MoE) models have become the dominant paradigm for large language models (LLMs), scaling from billions to trillions of parameters~\cite{kimiteam2026kimik3openfrontier,deepseekai2026deepseekv4highlyefficientmilliontoken,deepseek-v3}.
By selectively activating experts, MoE balances model capacity with per-token computation overhead. 

Consequently, Expert Parallelism (EP)~\cite{10.1145/3718958.3750506,deepep2025} has emerged as the primary deployment strategy for MoE models.
Under EP, experts are partitioned across multiple GPUs, tokens are dispatched to their routed experts for computation, and the resulting outputs are subsequently aggregated.
In distributed inference clusters, increasing the EP degree effectively reduces redundant expert weights, which leaves more GPU memory available for KV caches and runtime execution. 
As a result, real-world MoE deployments scale EP across tens or even hundreds of GPUs~\cite{deepseek-v3}.
Moreover, driven by continuous advances in interconnect and network fabrics (e.g., the CloudMatrix384 superpod~\cite{zuo2025servinglargelanguagemodels} and NVL72/144~\cite{wei2026ultraepunleashmoetraining}), the EP scale continues to expand.

\begin{figure}[!t]
  \centering
  \setlength{\belowcaptionskip}{-10pt}
  \incfig{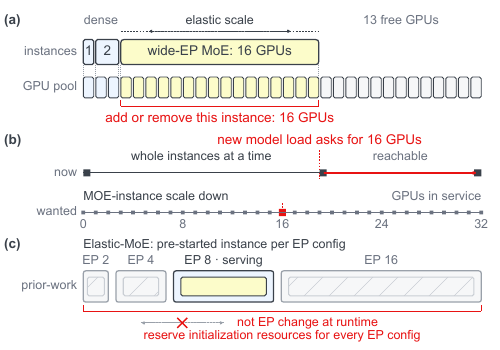}
  \caption{Fine-grained EP scaling. (a) A wide-EP instance occupies tens of GPUs, so
  instance-level scaling moves resources in units far coarser than
  demand changes. (b) A new model need not evict the MoE
  instance: the instance can give up some GPUs and keep serving.
  (c) ElasticMoE~\cite{elasticmoe} pre-starts one instance per EP configuration but cannot
  change the EP degree of a running instance.}
  \label{fig:scaling-granularity}
\end{figure}

However, while a large EP degree reduces serving costs,
it compromises scaling elasticity.
Dynamic auto-scaling is essential for handling fluctuating compute demands in modern inference systems, e.g., per-model request spikes in Model-as-a-Service (MaaS)~\cite{xiao2026huaweicloudmodelasaservicecloudmatrix384}, dynamic sample completion and entry during offline RL rollouts~\cite{10.1145/3779212.3790172}, and the volatile availability of spot instances and opportunistic resources~\cite{10.1145/3620665.3640411}. 
Although extensive research~\cite{blitzscale,aegaeon,chiron,tokenscale,serverlessllm} has explored dynamic scaling for LLM inference, prior approaches fundamentally treat an entire model instance as the minimum scaling granularity.
While this instance-level scaling suffices for early dense models, where a single instance typically fits within a handful of GPUs, it becomes highly impractical for large-scale Expert Parallelism (EP) deployments spanning hundreds of GPUs and creates a severe dilemma.
Deploying with a large EP degree restricts scaling elasticity, inevitably leading to compute resource over-provisioning during low-demand periods. Conversely, opting for a smaller EP degree improves scaling elasticity, but incurs a heavy memory penalty due to the redundant replication of model weights.

To break this dilemma, this paper advocates for \textbf{\emph{intra-instance elasticity via dynamic EP scaling}}, 
which enables fine-grained elastic scaling by dynamically adjusting the EP degree at runtime.
During periods of surging compute demand, the system expands the EP degree to handle higher loads.
When demand decreases, \sys reduces the EP group size to release idle compute resources while keeping the serving instance active, thereby avoiding instance teardown and reconstruction.
Intra-instance EP scaling pushes scaling granularity down to the single-GPU level, ensuring that provisioned compute resources precisely match highly dynamic workload demands.

However, enabling dynamic EP scaling is non-trivial. 
An MoE inference instance operates as a stateful, tightly coupled system rather than a collection of independent workers. 
As a result, scaling the EP degree up/down requires reconstructing global cluster state.
Existing implementations trigger instance-wide execution stalls lasting from several seconds to tens of seconds, leaving GPU resources idle and causing severe Service Level Objective (SLO) violations for requests.

Our key insight is \emph{\textbf{partial state reuse}}. 
We divide serving state into persistent and reconfigurable components.
Reusing persistent structures, such as per-GPU runtimes and pre-negotiated
collective communicators, avoids expensive distributed re-initialization.
Asynchronous online reconfiguration updates the remaining state while serving
continues, leaving only a brief coordinated transition.

To this end, we present \sys, an elastic MoE inference system that operationalizes partial state reuse. Specifically, \sys addresses the following three challenges.

\textbf{Challenge-1 (Network): Immutable communication groups.} 
EP inference relies on Data Parallel (DP) collectives and EP All-to-All communication, both of which operate over pre-established communicators (e.g., NCCL groups~\cite{nccl-analysis}).
Under the Single Program, Multiple Data (SPMD) paradigm, these collective operations mandate that every rank enters the call before execution can proceed. 
Existing communication groups cannot dynamically accommodate changing membership. 
Every EP scale-up/down forces the cluster to tear down active communicators and perform distributed re-initialization for new ones, which takes several to tens of seconds and incurs a cluster-wide stop-the-world pause.

\textbf{Challenge-2 (Memory): Memory reclamation destroys execution contexts.}
Scaling down must aggressively release GPU memory. 
Although model weights and KV caches consume a large share of memory capacity, our empirical analysis reveals that transferring/reloading them is not the primary latency bottleneck. Instead, rebuilding components of the execution context, such as computation graph~\cite{cuda-graph} buffers, kernel workspaces, and communication buffers, accounts for the vast majority of scaling latency.
Purging these runtime states destroys persistent execution structures and forces costly network handshakes and graph recapture during scale-up.

\textbf{Challenge-3 (Runtime): Safe worker retention and reuse. }
To eliminate GPU worker creation and initialization overheads during scale-up, the system retains scaled-down workers so they can be recycled. 
A target worker must stop GPU computation and remain alive, while active
workers continue serving outside a brief coordinated transition.
However, a worker's execution is tightly coupled with the rest of the cluster: it computes attention for its local requests and executes expert layers for tokens dispatched from remote ranks. 
Even if a worker drains its local requests, remote GPUs may still route tokens to its hosted experts. 
The system must therefore toggle a worker's participation safely without
discarding its initialized state.

To solve the challenges, we propose the following designs.

\textbf{Design-1: Reusing communicators through flexible group splitting.}
\sys creates a parent NCCL communicator over the maximum worker set and
retains its rank identities and bootstrap state. Each resize derives a subgroup
containing only active workers, so collectives do not wait for inactive ranks.
For expert dispatch and combine, a NIXL mask hides the same inactive ranks
while connections among active peers remain in place. Switching the subgroup
and mask together changes who communicates without rebuilding the distributed
instance. Because every subgroup comes from the parent, later scale-up can
reintroduce previously inactive workers.

\textbf{Design-2: Reusing GPU context with fine-grained offloading.}
\sys reclaims HBM from an inactive worker while preserving the runtime objects
that are expensive to create. CUDA virtual memory lets it retain the device
addresses of weights, KV-cache allocations, and graph buffers while releasing
the physical pages behind them. Contents follow their use: data that must
survive is backed up, replicated weights are fetched from active workers, and
request-specific KV state is moved or reconstructed. Graph buffers return to
the addresses recorded in captured CUDA Graphs, allowing replay without
recapture. Communication buffers are reclaimed only after remote access has
stopped, so their release does not invalidate the retained communication state.

\textbf{Design-3: Safe worker retention and reuse through coordinated sleep.}
\sys drains or moves requests off a target worker while making its experts
available on the remaining workers. At a safe point, it switches request
routing, expert placement, and communication membership together. After
in-flight work finishes, the target skips GPU execution but keeps its process,
executor, and CPU control threads alive; peers no longer send it tokens. On
scale-up, the same worker rejoins only after its memory and connections are
restored, avoiding reinitialization.

We implement \sys on vLLM and evaluate it across EP sizes from 2 to 16. Across
six transitions on each measured model, \sys scales in 4.2--13.4\,s, achieving
geometric-mean speedups of 7.0$\times$ over vLLM Elastic EP and 22.0$\times$
over restart. \sys also preserves steady-state throughput and enables colocated services to release and
reassign GPUs without instance reconstruction.

%% file: 03-background.tex
\begin{figure*}[!t]
  \centering
  \setlength{\belowcaptionskip}{-10pt}
  \scalebox{1}[0.75]{%
    \includegraphics[width=\textwidth]{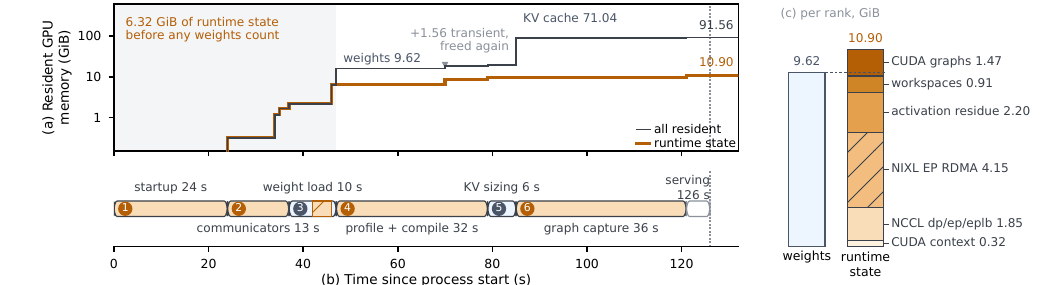}%
  }
  \caption{Bringing one DeepSeek-V2-Lite rank online
  ($\mathit{DP}{=}8$, $\mathit{EP}{=}8$). (a) Runtime state reaches
  6.32\,GiB before weights are accounted for. (b) The rank reaches steady state
  after 126\,s in six phases, of which only weight load moves model data.
  (c) The resulting 10.90\,GiB of runtime state exceeds the 9.62\,GiB of model
  weights on the rank.}
  \label{fig:boot-timeline}
\end{figure*}

\section{Characterizing the Cost of MoE Scaling}
\label{sec:bg}

Large-scale MoE inference requires dynamically provisioning GPU resources to match fluctuating traffic demands while maximizing compute utilization. However, Expert Parallelism (EP) fundamentally dictates both how GPU memory is partitioned and how flexibly resources can be reclaimed.

\subsection{MoE Serving and Elasticity Requirement}
\label{sec:bg:moe}

\paragraph{Expert parallelism.}
In MoE architectures, each token is dynamically routed to its top-$k$ experts. Because expert parameters dominate modern model footprints (e.g., 97\% of all parameters in DeepSeek-V3~\cite{deepseek-v3}), Expert Parallelism (EP) partitions these expert weights across $EP$ ranks. Spreading experts reduces the per-GPU weight memory footprint, freeing HBM to accommodate larger KV caches. Attention is typically replicated across data-parallel (DP) workers or sharded under tensor parallelism (TP), with serving engines enforcing $\mathit{EP}=\mathit{DP}\times\mathit{TP}$~\cite{vllm}. Production clusters routinely deploy EP degrees in the hundreds~\cite{deepseek-v3,deepseek-inference}, driven further by advancements in high-speed interconnect fabrics~\cite{zuo2025servinglargelanguagemodels,wei2026ultraepunleashmoetraining}.

\paragraph{Demand moves faster than instances do.}
Real-world LLM inference workloads exhibit extreme short-term traffic volatility. In the two-month BurstGPT trace~\cite{burstgpt} (1.4M requests), 59\% of ten-second windows experience no incoming requests, while busy periods exhibit a 25$\times$ swing between median and 99th-percentile arrival rates. Even in denser production environments like the Azure conversation trace~\cite{azurellmtrace,dynamollm} (27M requests over one week), traffic fluctuates by 1.7$\times$ with equally transient spikes.

Facing such fluctuations, cluster operators are forced into a severe dilemma: provision for peak traffic and suffer severe idle overheads during troughs, or provision for median traffic and violate Service Level Objectives (SLOs) during bursts. This inefficiency is unsustainable in multi-tenant clusters where MaaS platforms, RL rollouts, and spot workloads contend for shared GPUs~\cite{aegaeon,muxserve,alpaserve,xiao2026huaweicloudmodelasaservicecloudmatrix384,10.1145/3779212.3790172,10.1145/3620665.3640411}. Despite extensive efforts~\cite{chiron,tokenscale,enova,taming-chaos,hfx,serverlessllm,lambdascale,hydraserve,blitzscale}, existing frameworks operate strictly at the coarse granularity of whole instances, preloaded configurations, or full replicas. Crucially, none permit dynamically reconfiguring the active EP degree of a live, running instance.

\paragraph{The scaling must be fine, fast, and reversible.}
Effective intra-instance scaling must satisfy three requirements:
\begin{itemize}[leftmargin=*]
  \item \textbf{Fine-grained}: As shown in Figure~\ref{fig:scaling-granularity}, instance-level scaling operates at a granularity far too coarse (dozens of GPUs) for demand shifts.
  \item \textbf{Fast}: Existing intra-instance resizing takes 36.4--92.0\,s (\S\ref{sec:bg:runtime}), which is significantly longer than the median duration of the traffic excursions it is meant to handle.
  \item \textbf{Reversible}: Because traffic bursts are transient, resources added during a spike must be promptly reclaimed without incurring heavy tear-down and rebuild penalties.
\end{itemize}

\subsection{The Cost of MoE EP Resizing}
\label{sec:bg:runtime}

\paragraph{Existing designs resize whole instances.}
Current LLM serving systems manage capacity changes through coarse-grained instance manipulation or static pre-allocation. A naive restart tears down the entire instance and initializes a new one from scratch. vLLM's Elastic EP avoids process restarts but still reconstructs the distributed runtime—re-creating Ray workers, rebuilding process groups, and recapturing CUDA graphs~\cite{vllm-eep-blog}. Conversely, preloading approaches keep all parallel configurations resident simultaneously, paying exorbitant memory penalties~\cite{blitzscale,serverlessllm,hydraserve,lambdascale}. Runtimes with shadow modes preserve processes and execution graphs but freeze the parallel topology. Similarly, fault-tolerance mechanisms (e.g., SGLang rank-failure recovery~\cite{sglang-rankfail}) mask failed peers and remap experts without yielding physical memory back to the GPU pool. While re-parallelization and migration have been studied for full instances~\cite{10.1145/3620665.3640411,llumnix} and systems like ElasticMoE require warmup periods during scale-up~\cite{elasticmoe}, no existing system can dynamically reconfigure the active EP degree of a live instance while returning individual GPUs to the shared pool.

\paragraph{A resize costs a rebuild.}
Figure~\ref{fig:boot-timeline} quantifies the latency breakdown of an instance rebuild, taking 126\,s from process launch to steady-state serving across six distinct phases: 24\,s startup, 13\,s communicator construction, 10\,s weight loading, 32\,s operator profiling and compilation, 6\,s KV sizing, and 36\,s CUDA graph capture. Crucially, loading model weights takes only 10\,s; the remaining 116\,s is consumed by building the execution environment. In fact, 6.32\,GiB of HBM is already consumed before model weights are loaded at $t=47\text{\,s}$. At steady state, runtime context overhead reaches 10.90\,GiB, exceeding the 9.62\,GiB of model weights, with full boot times spanning 156--267\,s across configurations. Under a closed-loop $EP\,8 \rightarrow 4 \rightarrow 8$ scaling cycle, vLLM Elastic EP requires 47\,s to shrink and 75\,s to grow, incurring 31\,s and 44\,s of complete service unavailability, respectively; a full process restart inflates service downtime to 141\,s and 149\,s (\S\ref{sec:eval:roundtrip}). Furthermore, service downtime is compounded by post-scaling load balancing (e.g., EPLB~\cite{vllm-eplb}), which causes sustained throughput degradation before converging.

Analyzing the state destroyed during a resize (Table~\ref{tab:resize-state}) reveals that the overhead stems primarily from components invariant to the EP degree itself. Worker processes, compiled kernels, CUDA contexts, communication infrastructure, and captured graphs depend strictly on the model architecture, device, and maximum world size rather than the immediate EP degree. Memory allocations follow a similar trend: because NIXL transfer buffers are dimensioned for the maximum world size, an instance capable of scaling up to eight ranks allocates memory for eight ranks from startup, much of which is allocated outside the serving engine's managed pools by driver-level libraries. What the EP degree truly dictates is merely a minimal set of dynamic state: active rank participation, expert-to-node placement mappings, physical page residency, and peer connectivity.

\paragraph{Three rigid bindings force the rebuild.}
Current inference runtimes enforce three structural bindings that make dynamic EP scaling impossible without a full rebuild:
\begin{itemize}[leftmargin=*]
  \item \textbf{Immutable Communicator Binding}: Collective communication groups are constructed for fixed rank sets. Under standard SPMD collective semantics, every member rank must synchronously enter a call, meaning dropping or adding a rank forces the destruction and distributed re-initialization of the process group—unless subcommunicators are derived from a persistent parent group spanning the maximum rank set~\cite{nccl-analysis}.\label{sec:bg:comm}
  \item \textbf{Static Memory \& Address Binding}: Captured CUDA graphs, operator kernel arguments, and registered RDMA memory regions tightly bind to hardcoded virtual memory addresses. Deallocating or resizing a buffer invalidates the execution path even when data shapes remain unchanged—unless virtual address spaces are decoupled from underlying physical page allocations.\label{sec:bg:vmm}
  \item \textbf{Tightly Coupled Worker Execution}: An inference worker concurrently manages local request state, KV caches, and hosted experts serving remote token dispatches. A worker cannot independently pause or exit without triggering remote hangs—unless remote peers dynamically mask the rank and remap expert routes, similar to fault-recovery mechanisms~\cite{sglang-rankfail}.
\end{itemize}

While each binding stems from software implementation conventions rather than hardware constraints, no deployed system systematically breaks all three bindings to enable low-latency intra-instance EP scaling on live workloads.

\begin{table}[t]
  \centering
  \tabsetup
  \caption{What a resize takes apart, itemized by state. Costs are the phases
  measured in Figure~\ref{fig:boot-timeline}; footprints are steady-state
  values on the same rank. Only the last row is a function of the active EP
  degree.}
  \label{tab:resize-state}
  \begin{tabular}{@{}L{0.25\linewidth}L{0.155\linewidth}C{0.13\linewidth}L{0.165\linewidth}L{0.165\linewidth}@{}}
    \toprule
    \colhd{State} & \colhd{Cost} & \colhd{EP-dep.} & \colhd{Today} & \colhd{Reused} \\
    \midrule
    Worker process, executor, kernels & 24\,s startup & no & destroyed & retained \\
    Communication domains, RDMA buffers & 13\,s setup & members only & rebuilt & subdomain derived \\
    CUDA context & 0.32\,GiB & no & destroyed & retained \\
    CUDA Graphs & 36\,s, 1.47\,GiB & no & recaptured & replayed \\
    Weights, KV cache, workspaces & 9.62 / 71.04\,GiB & sizes only & reloaded & pages refilled \\
    Expert placement, request routing & \naa & yes & rebuilt & rewritten in place \\
    \bottomrule
  \end{tabular}
\end{table}

%% file: 04-overview.tex
\section{FastEEP Overview}
\label{sec:motivation}

Given a running MoE serving instance, \sys changes its EP size quickly while
minimizing disruption to request processing.
Scaling down releases GPUs for other workloads; scaling up restores GPU
capacity when throughput and
KV-cache demand increases.

\begin{figure*}[!t]
  \centering
  \setlength{\belowcaptionskip}{-10pt}
  \includegraphics[width=\textwidth]{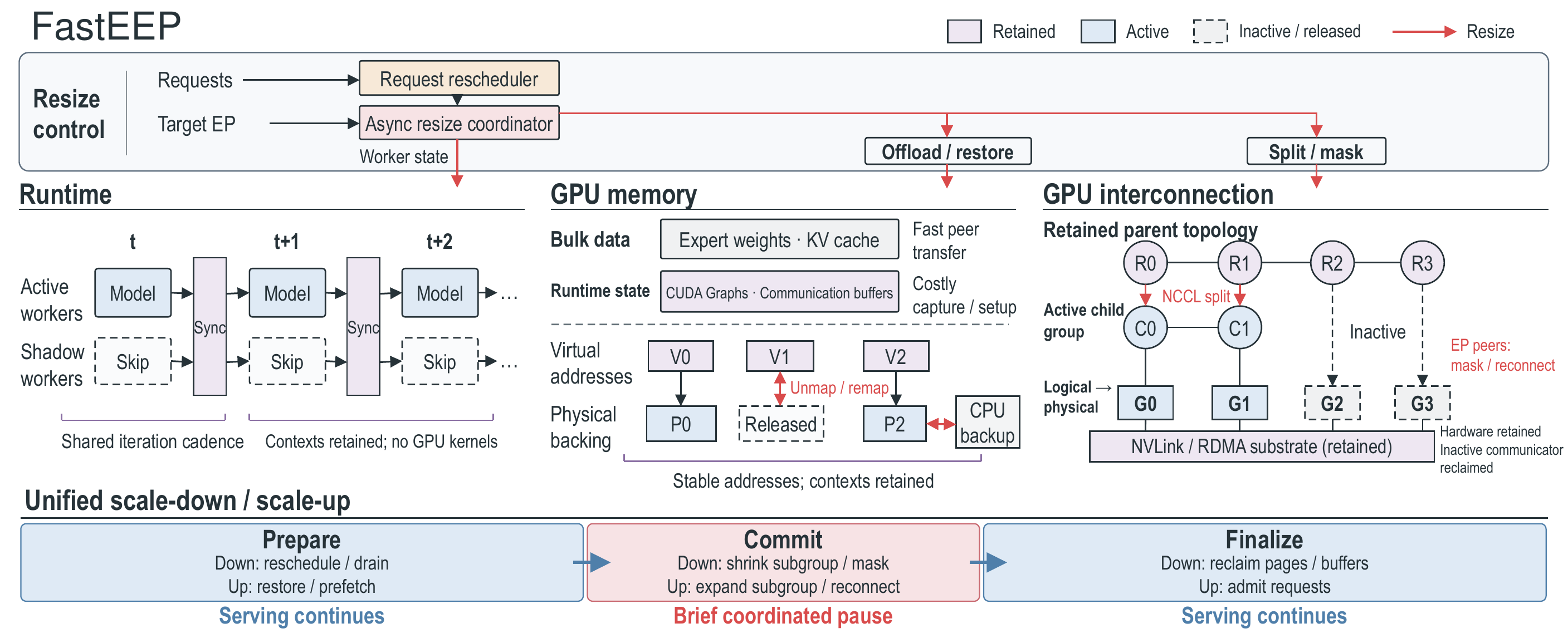}
  \caption{Architecture of \sys. A maximum-size instance retains worker
  execution contexts, a parent communication domain, and virtual address
  ranges. Resizing changes the active worker set and physical-memory residency
  without reconstructing the instance.}
  \label{fig:flashepscale-architecture}
\end{figure*}

\subsection{Existing Designs}
\label{sec:motivation:breakdown}

Existing systems expose two approaches to changing the EP size of a serving
instance.

\paragraph{Preloaded-instance switching.}
One approach pre-initializes several instances of the same model with different
parallel configurations.
Inactive instances offload their weights and KV cache to host memory, then
restore them when selected.
This avoids a full cold start, but each configuration still retains its
processes, CUDA Graphs, and communication state.
For MoE serving, that residual footprint is substantial: communication and
driver allocations alone consume several GiB per GPU, while the full runtime state can exceed
10\,GiB (\S\ref{sec:bg:runtime}).
Moreover, preloading can cover only a fixed set of configurations, trading
memory consumption for limited scaling flexibility.

\paragraph{Intra-instance reconstruction.}
The alternative reconstructs the running instance at the target size.
vLLM Elastic EP~\cite{vllm-eep-blog} follows this approach: it changes the Ray
worker set, rebuilds communication groups, and captures new CUDA Graphs.
SGLang provides a related mechanism for rank-failure
recovery~\cite{sglang-rankfail}, rather than deliberate resource scaling.
Reconstruction supports more configurations than preloading, but repeats
expensive initialization on every transition.
In our experiments, an Elastic EP resize takes 36.4--92.0\,s
(\S\ref{sec:eval:endtoend}), too long to track short-term workload changes.

\subsection{Challenges}
\label{sec:motivation:state}

\paragraph{Reconfiguring a fixed communication fabric.}
Wide-EP inference couples two communication paths with different semantics:
DP collectives synchronize attention workers, whereas EP All-to-All,
implemented as point-to-point transfers, dispatches tokens to remote experts
and returns their outputs.
Their rank membership, transport endpoints, topology, channels, and registered
buffers are established during initialization (\S\ref{sec:bg:comm}) and are
assumed to remain fixed during execution. Under SPMD collective semantics, a
single inactive rank can prevent the remaining ranks from making progress;
conversely, stale EP routing may direct remote accesses to buffers that have
already been reclaimed. An EP resize must therefore switch both communication
paths to one consistent active-rank view. Reconstructing that view from
scratch requires distributed bootstrap and topology setup---close to 20\,s for
an eight-GPU group in our measurements---and imposes a cluster-wide stall.

\paragraph{Reclaiming HBM without invalidating runtime state.}
Scaling down must release not only model weights and KV cache, but also the
CUDA Graph storage, operator workspaces, and communication buffers that account
for a substantial fraction of each worker's HBM footprint. These allocations
are not interchangeable byte arrays: captured CUDA Graph executables embed
their device addresses, while communication libraries retain registrations and
remote-access metadata for their buffers (\S\ref{sec:bg:vmm}). Conventional
deallocation destroys these address and registration invariants, so scale-up
must rebuild communicators and repeat graph warm-up and capture even if tensor
shapes are unchanged. In our evaluated configurations, graph recapture alone
takes 25--82\,s. The memory challenge is thus to reclaim physical HBM without
discarding the logical allocation identity required to reuse the execution
context.

\paragraph{Quiescing workers without stopping service.}
Avoiding worker reconstruction requires a scaled-down worker to remain alive
but quiescent while the rest of the instance continues serving. Reaching this
state is difficult because each worker simultaneously owns local requests and
their KV cache, participates in DP synchronization, and executes experts for
tokens routed from remote ranks. Draining its local request queue therefore
does not eliminate remote expert traffic or outstanding communication. A safe
transition must jointly update request ownership, KV-cache placement, expert
placement, token-routing tables, communication membership, and in-flight
operations before any compute or memory resource is released. Otherwise, the
resize can lose request state, invoke a collective with mismatched membership,
or access reclaimed memory. This cross-subsystem transition must complete with
only a brief coordinated pause rather than stopping the entire instance for
worker teardown and reconstruction.

\subsection{System Overview}
\label{sec:motivation:overview}

\sys replaces reconstruction with \emph{partial reuse}: it initializes the instance once, retains expensive long-lived state, and changes
only the resources that participate in the active configuration.

At the memory layer, tagged CUDA virtual-memory pools preserve virtual
addresses while selected physical pages are offloaded or released.
Weights can be restored from host memory or active peers, expired KV-cache
pages can be discarded, and CUDA Graph storage can be restored without
recapture.
At the communication layer, \sys derives active NCCL subgroups from a
persistent parent group\footnote{The parent group remains logically
persistent, but after a resize, \sys suspends it and releases the GPU memory
occupied by its data buffers, restoring them when the group is needed again.}
and reconnects or masks NIXL peers instead of rebuilding the entire
communication domain.

At the execution layer, \sys removes target workers from request routing,
drains or re-issues their in-flight requests, remaps required experts, and
masks inactive ranks from expert dispatch.
The worker processes, executors, and compiled kernels remain alive for rapid
reactivation.

A resize is executed as a staged transaction: prepare resources concurrently,
commit the new active configuration at a short global quiescence point, and
reclaim idle resources after request processing resumes.
Together, these mechanisms preserve correctness while reducing scaling latency
by 4.5--11.7$\times$ relative to vLLM Elastic EP
(\S\ref{sec:eval:endtoend}).

%% file: 05-design.tex

\section{FastEEP Design}
\label{sec:design}
\label{sec:overview}

The three challenges above arise from a common cause: existing systems bind
active resources to initialized runtime state within one lifecycle. Changing
the EP size therefore destroys not only GPU resources, but also communication
domains, device addresses, CUDA Graphs, and worker contexts, all of which must
be rebuilt during scale-up.

\sys addresses this problem through \emph{partial state reuse}, separating
stable structural state from active state that changes with the EP size. It
first decouples the physical communication network from logical participants,
allowing different EP sizes to share one substrate. It then decouples virtual
addresses from physical HBM, allowing shadow workers to release memory while
preserving their execution contexts. Finally, the runtime combines changes to
communication, memory, request routing, and expert placement into one
consistent worker-state transition.
\label{sec:overview:premise}

\subsection{Reusing the Communication Network}
\label{sec:comm}

MoE inference uses collectives for synchronization and aggregation and
All-to-All communication for token dispatch and combine. Point-to-point peers
can be controlled independently, but collective participants are fixed in a
group. Rebuilding that group requires bootstrap, topology discovery, connection
setup, and buffer allocation. \sys instead retains a global group over all
workers. It suspends this group and releases its resources when idle, then
reuses its context and control-plane state to derive a group for the current
active workers.

\begin{figure}[t]
  \centering
  \includegraphics[width=\columnwidth]{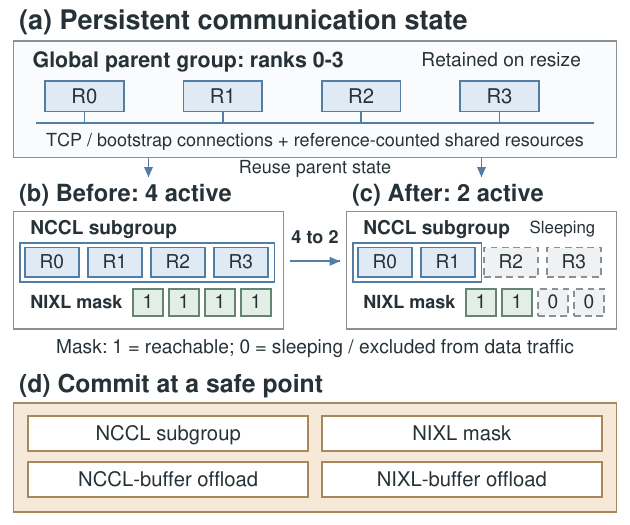}
  \caption{Reusing communication state across a resize. A persistent parent
  domain retains the global rank space and bootstrap context. Scaling from four
  to two workers changes the NCCL subgroup and NIXL reachability mask. These
  views, the expert mapping, and request routing are committed together.}
  \label{fig:communication-reuse}
\end{figure}

\paragraph{All-to-All communication: dynamically changing point-to-point
reachability.}
MoE dispatch and combine logically form an All-to-All operation. Resizing
changes both active ranks and expert placement, so each token must reach the
rank currently hosting its expert and return to its source under the same
active view.

Point-to-point transfers naturally support dynamic membership because each
peer has an independent lifecycle. NIXL EP uses UCX across nodes and separates
remote-metadata exchange from endpoint creation~\cite{nixl}. It can cache peer
metadata and create endpoints lazily or explicitly. Adding or removing a rank
therefore affects only its endpoints, leaving other connections intact.

\sys retains the maximum-size \emph{logical rank space} and represents visibility in
a device-side mask read by dispatch and combine kernels. During scale-down, it
masks a rank after replicas of its experts become available. During scale-up,
it unmasks the rank only after its buffers and endpoints are ready. Membership
thus changes without rebuilding the All-to-All domain.
\label{sec:exec:mask}

\paragraph{Collective communication: retaining the parent domain and switching
the active subgroup.}
NCCL communicator membership, however, is a synchronization contract: every member must
enter each collective. As long as an inactive rank remains a member, all other
ranks must wait for it. Masking inactive ranks in P2P operators does not resolve this issue 
because the inference communicator must include only the active ranks.

\sys therefore uses a two-level communication domain. At startup, it creates a
parent domain spanning the maximum worker set and retains its rank space,
bootstrap relationships, and control-plane state throughout the lifetime of
the instance (less than 300\,MB in total).
Its data-plane resources can be selectively offloaded to host memory during scaling.
Each resize derives from this parent a
working subgroup containing only active ranks. Scale-down installs a smaller
subgroup and scale-up a larger one. Subgroups are always derived from the
parent, so contraction never removes relationships needed for expansion.

NCCL communicator split implements this structure~\cite{nccl-analysis}. Active ranks
receive the same color and form a subcommunicator, and inactive ranks remain
outside it. Reusing the parent's global identities and bootstrap context avoids
worker restart and reconstruction of the distributed environment.
\label{sec:comm:split}

\paragraph{Unifying both communication paths with one active set.}
The NCCL subgroup and NIXL mask represent the same active set. The former makes
DP collectives wait only for active workers; the latter restricts EP dispatch
and combine to them. At one safe point, \sys publishes both views with the
logical-expert-to-physical-slot mapping, making each rank visible to either all
three structures or none.
\label{sec:comm:dual-track}

\subsection{Fine-Grained GPU Memory Management}
\label{sec:mem}

Leaving distributed execution does not release a worker's GPU memory. Besides
weights and KV cache, it holds CUDA Graph buffers, operator workspaces,
communication buffers, and runtime caches whose setup dominates startup cost.
\sys therefore preserves their logical identities and execution contexts while
offloading contents or unmapping physical pages. A shadow worker (see \S\ref{sec:exec}) can reclaim
HBM and later restore this state in place.

\begin{figure}[t]
  \centering
  \includegraphics[width=\columnwidth]{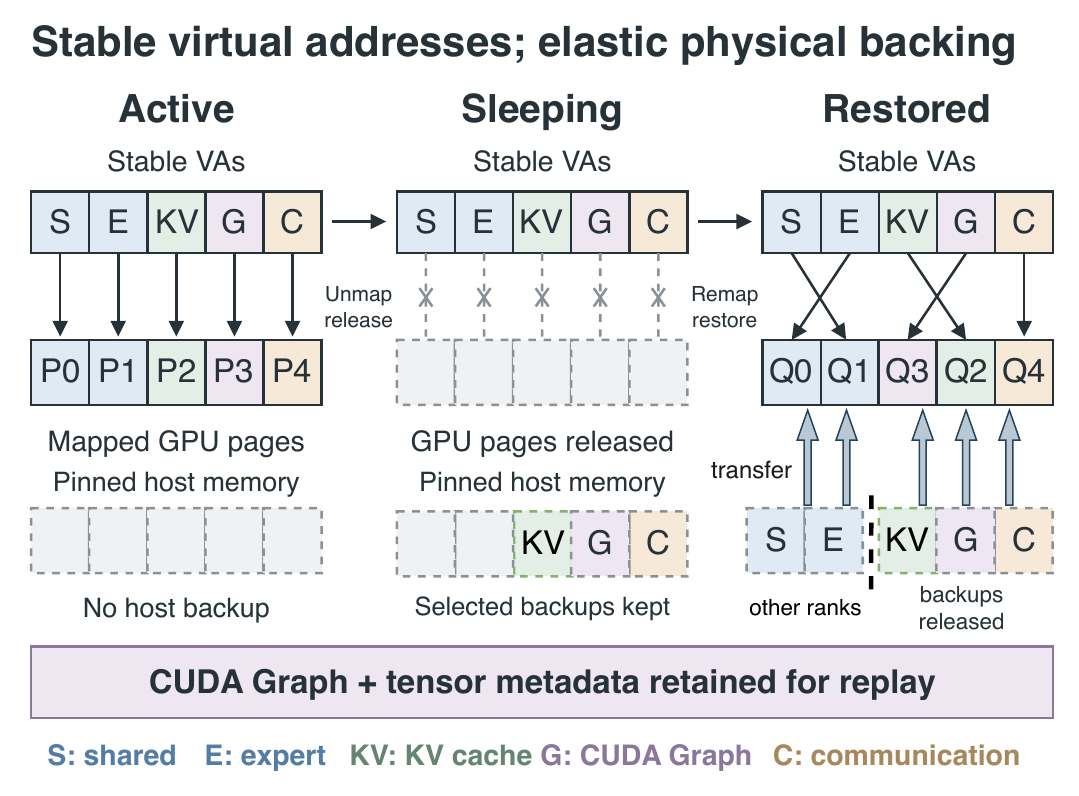}
  \caption{Address-stable memory management. A shadow worker retains virtual
  address ranges while selected physical pages are released. On wake-up, new
  pages are mapped at the same addresses and restored according to allocation
  semantics.}
  \label{fig:memory-management}
\end{figure}

\paragraph{Preserving virtual addresses while changing physical mappings.}
Using CUDA Virtual Memory Management, \sys creates separate
CuMem pools for shared weights, expert weights, KV cache, and CUDA Graph
buffers. Scale-down unmaps and releases physical pages while retaining virtual
addresses, tensor metadata, and object references. Scale-up maps new pages at
the same addresses.

Stable addresses preserve kernel arguments, tensor pointers, and allocator
objects, allowing wake-up without rebuilding runtime structures. Irreplaceable
contents are backed up before unmapping; data available from another worker or
cheap to reconstruct is discarded. This trades host-memory use for recovery
latency without sacrificing address reuse.
\label{sec:mem:semantics}

\paragraph{Restoring weights and KV cache according to their semantics.}
Under EP, non-expert weights such as attention weights are replicated. A
shadow worker can discard them and later restore them from an active rank over
NVLink at their original addresses. Expert weights use the same device-to-device
path. The control plane computes the target placement and source--destination
rank pairs so that every logical expert has a replica in the active set. It
batches transfers by tensor or expert, while the data plane performs
asynchronous peer copies on a separate CUDA stream.

For request-specific KV cache, \sys chooses among three policies for each block:
discard and recompute it from the prompt and generated tokens after
rescheduling; migrate it asynchronously to an active rank over NVLink; or swap
it to pinned host memory and reload it later. This balances recovery latency,
interconnect bandwidth, and host-memory use without retaining a shadow
worker's complete KV cache.
\label{sec:mem:offload}

\paragraph{Offloading CUDA Graph buffers.}
CUDA Graph capture fixes kernel arguments and device pointers in the graph
executable. A reallocated buffer therefore invalidates replay if its address
changes, even when its shape and contents do not. Repeating warm-up and capture
also adds substantial scale-up latency.

\sys routes capture-time allocations to a dedicated CuMem graph pool.
During the sleep transition, \sys backs up required contents and unmaps
physical pages while preserving virtual addresses; on wake-up, it remaps
pages and restores data at the same locations. The
graph's pointers, arguments, and topology remain unchanged, enabling direct
replay without recapture. Thus, \sys preserves the graph execution context,
not its original physical pages.
\label{sec:mem:graph}

\paragraph{Communication-buffer reachability.}
Communication buffers may remain accessible to remote peers or in-flight
kernels. Releasing one locally can therefore redirect remote writes into
reallocated pages. \sys uses collective-group switching and point-to-point
visibility as safe lifetime boundaries for these buffers.

For collectives, \sys first installs the new active subgroup and drains old
operations. It then offloads idle parent-group buffers to pinned host memory
and releases their HBM, retaining the group identity, rank information, and
bootstrap context. Before the next resize, it restores the buffers at their
original addresses and derives a new subgroup. The parent can thus persist
without keeping its full data plane in HBM.
\label{sec:comm:suspend}

For point-to-point communication, the NIXL mask controls reclamation. During
scale-down, the target is first removed from expert routing and the mask; after in-flight
traffic drains and endpoints disconnect, \sys reclaims its NIXL buffers.
Scale-up restores and registers buffers and establishes endpoints while the
rank remains invisible, then unmasks it. Both paths follow one rule: revoke
reachability before releasing memory, and restore memory before granting
reachability.
\label{sec:mem:comm-buffer}
\label{sec:mem:rdma-truth}

\subsection{Reusing Worker Execution Contexts}
\label{sec:exec}

The preceding mechanisms make communication membership switchable and
decouple memory identity from physical residency. The runtime combines them
into one consistent worker-state transition.

Partial reuse changes worker state rather than creating or destroying workers.
Scale-down stops target computation, removes the worker from the data plane,
releases network buffers, and offloads physical memory. Its process, executor,
compiled kernels, virtual addresses, CUDA Graph executables, and control state
remain intact. Scale-up restores memory and communication reachability and
re-enables computation, avoiding process startup, model initialization, graph
capture, and communication reconstruction.

\paragraph{The shadow worker state.}
A shadow worker retains its execution identity without consuming GPU compute resources. 
Its process, CPU threads, executor, virtual addresses and
control state persist. Physical memory for weights, KV cache, graph buffers,
and communication buffers is unmapped or offloaded, leaving only the
device state required for recovery. Wake-up restores these contents at their
original addresses.

The worker is excluded from the NCCL subgroup and hidden by the NIXL mask, so
it receives neither collectives nor dispatch and combine traffic. Its global
rank, bootstrap information, and control connections remain available for
recovery and later reconnection.

Before model execution, the runtime checks the active state. Shadow workers
bypass the forward pass and launch no attention, expert, or communication
kernels. Their CPU threads handle only control commands and barriers, yielding
GPU compute while remaining ready to wake at a coordinated transition point.

\paragraph{Algorithm-level contraction.}
In DP+EP, attention parallelizes requests along DP, while experts parallelize
tokens along EP. Each worker handles local requests and KV cache while serving
expert tokens from peers. Scale-down must close both paths: stopping request
scheduling alone does not stop remote expert traffic.

\begin{figure}[t]
  \centering
  \includegraphics[width=\columnwidth]{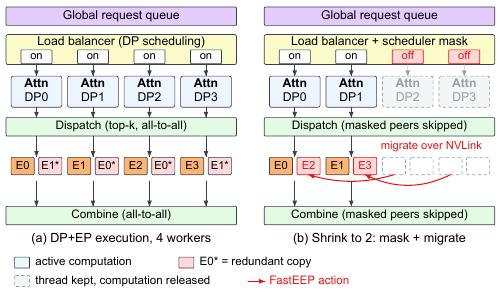}
  \caption{Contracting DP and EP execution during scale-down. The scheduler
  stops routing new requests to target workers and drains or migrates their
  current requests. Missing experts are copied into available slots on the
  remaining workers before the target ranks leave the EP data path.}
  \label{fig:dpep}
\end{figure}

On the request path, \sys marks the target as draining and removes it from the
active DP set. Existing requests drain or migrate. Once no active request or KV
state remains, the runtime skips its attention forward pass.

On the expert path, \sys recomputes the logical-expert-to-physical-slot mapping,
reuses replicas on active workers, and copies missing experts into free slots
over NVLink. Once every logical expert is available, it publishes the mapping
and NIXL mask together, removing targets from dispatch and combine. Available
expert slots therefore determine the minimum feasible active size,
$N_{\min}$.
\label{sec:overview:invariants}

Only after both paths contract can the target sleep and release compute,
communication, and HBM resources. Scale-up reverses this order: it restores
memory, expert placement, and communication before admitting the worker to the
active DP set.

\paragraph{Decoupling request and worker lifecycles.}
After a target begins draining, near-complete requests finish locally. \sys
moves the rest to active workers and, based on recovery cost, recomputes their
KV cache, migrates live blocks over NVLink, or stores them in pinned host
memory.

The scheduler unbinds a request only after it completes or obtains another
execution location. \sys can then reclaim the original KV cache and sleep the
worker, decoupling request continuity from worker-resource lifetime.

The control plane moves workers between active and inactive states, using one
active set as the commit boundary for request routing, expert placement, and
communication. A configuration becomes visible only after all required
resources are ready.

\textbf{Scale-down.}
The control plane inspects each rank's state, request queue, and KV-cache
occupancy to select targets. It stops new admission, then drains requests or
migrates their state. In parallel, it rearranges experts and transfers any
weights missing from the contracted placement. Once requests and experts no
longer depend on the targets, \sys offloads their weights, KV cache, and CUDA
Graph buffers and releases the HBM. It derives a new subgroup from the global
NCCL group, updates the NIXL mask and endpoints, and switches views at a safe
point. After old operations finish, it offloads unused global-group and inactive
rank buffers to the CPU, then marks the targets inactive.

\textbf{Scale-up.}
The control plane selects shadow workers to wake, remaps their physical
pages, and reloads CUDA Graph and communication buffers from the CPU. Through
the retained global communication relationships, it copies non-expert weights,
expert weights, and required KV blocks from active ranks to their original
addresses. It then derives the expanded NCCL subgroup, registers NIXL buffers,
and establishes peer endpoints. Once all connections are ready, the control
plane atomically publishes the expert mapping, communication view, and request
routing and marks the ranks active. They can now receive requests without
reinitializing their execution contexts.

%% file: 06-evaluation.tex

\section{Evaluation}
\label{sec:eval}

We evaluate the performance benefits of \sys under real
workloads, focusing on its ability to reduce the latency of scaling a running
service and the resulting interference with live requests. We also evaluate
the effectiveness of dynamic scaling in multi-model deployments.

\paragraph{Testbed.}
Experiments run on two GPU nodes.
Each node has two Intel Xeon Platinum 8575C CPUs and eight NVIDIA H20-3e GPUs (140 GiB HBM each). The GPUs are connected by 900 GB/s intra-node NVLink. Inter-node communication uses eight BlueField-3 integrated ConnectX-7 NICs (400 Gb/s per NIC).

\paragraph{Implementation.}
\sys is implemented on vLLM
\texttt{v.0.27.0}, Python 3.12, PyTorch 2.11.0+cu130,
NCCL 2.28.9, and NIXL 1.2.0. The implementation adds about 5K lines of
production code for elastic EP scaling, request transfer, NIXL rank masking,
expert remapping, and weight refilling. These components are integrated with
the vLLM scheduler and GPU worker without changing the MoE serving interface.

\begin{figure}[t]
  \centering
  \setlength{\belowcaptionskip}{-10pt}
  \includegraphics[width=\columnwidth]{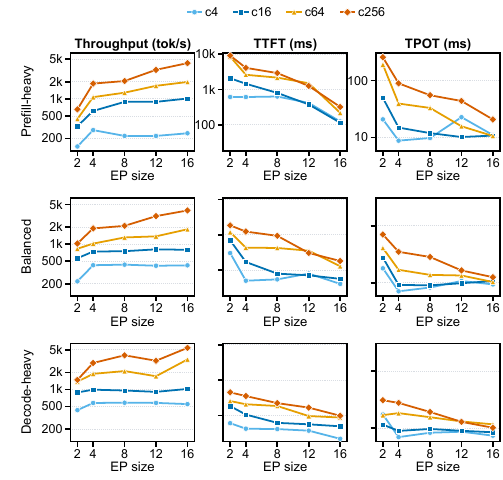}
  \caption{Serving performance (Qwen3-30B-A3B) across EP sizes for prefill-heavy, balanced,
  and decode-heavy requests (rows): output throughput, mean TTFT, and mean
  TPOT (columns).}
  \Description{Nine panels plot throughput and latency against EP size for
  three request shapes. Four curves in each panel represent concurrency 4,
  16, 64, and 256.}
  \label{fig:epsize}
\end{figure}

\begin{figure}[t]
  \centering
  \setlength{\belowcaptionskip}{-10pt}
  \includegraphics[width=\columnwidth]{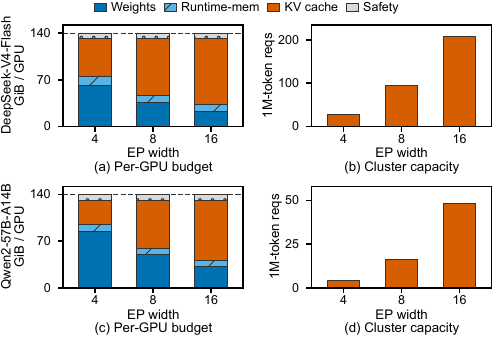}
  \caption{KV-cache capacity for H20 GPUs, \texttt{nixl\_ep}, and 128 redundant experts. Each model
  occupies one row: the left panel shows the per-GPU memory budget across EP
  widths, and the right panel shows realizable cluster-wide concurrency for
  $10^6$-token requests.}
  \label{fig:kvvsep}
\end{figure}

\begin{figure*}[t]
  \centering
  \includegraphics[width=\textwidth]{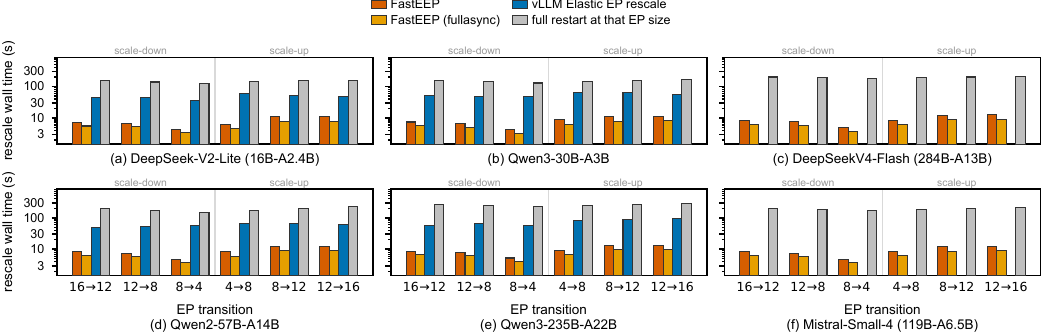}
  \caption{End-to-end EP scaling latency across six MoE models. vLLM Elastic
  EP implementation could not scale DeepSeek-V4-Flash or Mistral-Small-4
  (\texttt{NaN}). FullAsync refers to a mode in which the execution transition is
  synchronized with request processing.}
  \label{fig:exp2scale}
\end{figure*}

\paragraph{Baselines.}
For scaling comparisons, we evaluate three systems using the same software stack:
\begin{itemize}[leftmargin=*,labelindent=0pt,nosep]
  \item \emph{Restart.} It terminates the serving instance and
  relaunches vLLM with the target EP size.
  \item \emph{vLLM Elastic EP baseline~\cite{vllm-eep-blog}.} It creates or removes
  Ray engines and reinitializes distributed communication.
  \item \emph{\sys.} It leverages the elastic scaling mechanisms developed in this work for GPU communication, GPU memory management, and worker execution, thereby enabling efficient intra-instance scaling.
\end{itemize}

\paragraph{Models and workloads.}
Our scaling study covers six MoE models, including the
DeepSeek-V2-Lite~\cite{deepseek-v2}, Qwen3-30B-A3B-Instruct-2507~\cite{qwen3}, DeepSeek-V4-Flash~\cite{deepseekai2026deepseekv4highlyefficientmilliontoken},
Qwen2-57B-A14B~\cite{qwen2}, Qwen3-235B-A22B~\cite{qwen3}, and
Mistral-Small-4-119B-2603~\cite{mistral-small-4}.
We use two classes of workloads. To isolate the throughput characteristics of
EP, we use three fixed request shapes: prefill-heavy (4096 input / 128 output),
balanced (1024 / 1024), and decode-heavy (256 / 2048). To evaluate EP scaling
latency under realistic serving conditions, we use real-world requests sampled
from ShareGPT~\cite{sharegpt}. We evaluate EP sizes
$\{2,4,8,12,16\}$, covering both intra-node and cross-node scaling.

\subsection{Performance and KV-Cache Capacity across EP Sizes}
\label{sec:eval:epsize}

Figure~\ref{fig:epsize} compares EP trends in throughput and latency across
concurrency levels and request shapes.

\textbf{The best EP size depends on load and request shape.}
At concurrency 4, increasing EP from 4 to 16 yields 0.88$\times$,
0.99$\times$, and 0.95$\times$ the EP-4 throughput for prefill-heavy,
balanced, and decode-heavy requests, respectively, indicating that a wider EP
is underutilized at low load. At concurrency 256, the same change improves
throughput by 2.30$\times$, 2.08$\times$, and 1.85$\times$. Prefill-heavy
requests benefit most, while decode-heavy requests exhibit a flatter scaling
trend. Thus, no single EP size is optimal: \sys can select the smallest EP
that satisfies workload-specific TTFT and TPOT SLOs, scaling up as pressure
increases and releasing GPUs when a smaller EP is sufficient.

\textbf{EP size also controls runtime KV-cache capacity.}
As shown in Figure~\ref{fig:kvvsep}, increasing EP expands the available KV
cache, which is especially valuable
for high-concurrency, long-context and agentic workloads. \sys can therefore adapt EP to the
workload type, meeting memory demand while improving GPU memory utilization.

\subsection{End-to-End Scaling Latency}
\label{sec:eval:endtoend}
\label{sec:eval:measured}

We evaluate six scale-up and scale-down transitions among
$\{4,8,12,16\}$. Figure~\ref{fig:exp2scale} reports results for six
models.
\textbf{\sys consistently reduces scaling latency across models and
transitions.}
Across all models and six transitions, \sys completes scaling in
4.2--13.4\,s, while its FullAsync variant takes 3.2--9.8\,s. In
comparison, vLLM Elastic EP takes 36.4--92.0\,s, and a full restart takes 126--293\,s across all six models. Across
all model-transition pairs for which Elastic EP is available, \sys achieves a
4.5$\times$--11.7$\times$ speedup, with a 7.0$\times$ geometric mean; the
FullAsync variant increases the geometric-mean speedup to 9.3$\times$. Across
all model-transition pairs, the geometric-mean speedups over restart are
22.0$\times$ and 29.3$\times$ for \sys and its FullAsync variant,
respectively. Scale-up is generally more expensive than scale-down, but \sys
is faster than both baselines for every available comparison, consistent with
the benefit of reusing workers, communication state, and CUDA Graphs across
model sizes and transition directions. The official vLLM Elastic EP
implementation could not successfully run DeepSeek-V4-Flash and
Mistral-Small-4\footnote{The \texttt{BatchedMarlinExperts} implementation does not support Elastic EP.}; therefore, we omit Elastic EP results for these two models.

\begin{figure*}[t]
  \centering
  \includegraphics[width=\textwidth]{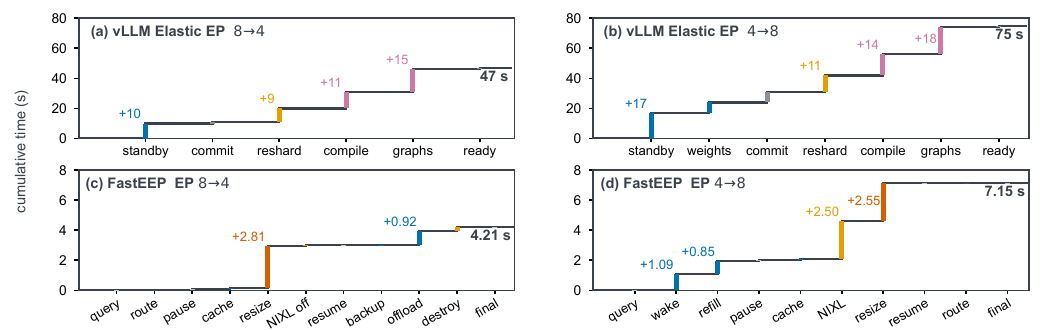}
  \caption{Cumulative time across EP scaling phases for vLLM Elastic EP
  (a, b) and \sys (c, d), in the $8\rightarrow4$ and $4\rightarrow8$
  directions.}
  \label{fig:exp5cum}
\end{figure*}

\begin{figure*}[!t]
  \centering
  \setlength{\belowcaptionskip}{-10pt}
  \includegraphics[width=\textwidth]{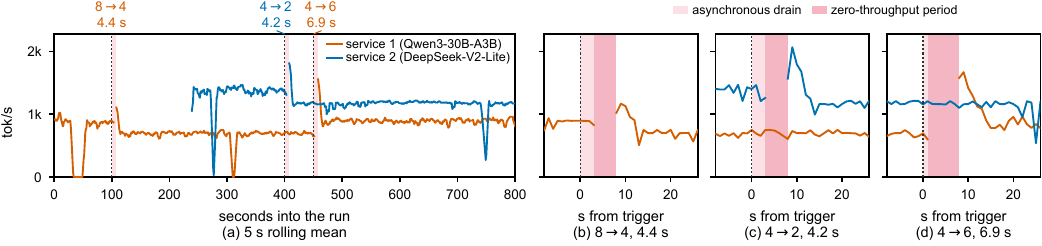}
  \caption{Resource-sharing trace for two colocated services: 5\,s rolling
  mean in panel~(a), raw 1\,s samples in panels~(b)--(d). Each service uses
  GPUs released by the other's scale-down.}
 \label{fig:multi}
\end{figure*}

\textbf{Expert re-layout and state restoration dominate the scaling path.}
Figure~\ref{fig:exp5cum} compares the phase breakdowns of vLLM Elastic EP
(a, b) and \sys (c, d) for the EP $8\!\rightarrow\!4$ scale-down and
$4\!\rightarrow\!8$ scale-up of Qwen3-30B. In the plotted vLLM breakdown,
scale-down totals 47\,s: standby setup takes 10\,s, expert resharding 9\,s,
compilation and warmup 11\,s, and graph capture 15\,s. Scale-up totals
75\,s, including 17\,s for standby setup, 7\,s for weight and KV
synchronization, 11\,s for expert resharding, 14\,s for compilation and
warmup, and 18\,s for graph capture.

For \sys, scale-down totals 4.21\,s. The \texttt{resize} phase, which
installs the new expert layout and communication view, takes 2.81\,s;
offloading the shadow workers' buffers takes 0.92\,s and destroying their
NIXL buffers takes 0.24\,s. Scale-up totals 7.15\,s. Waking workers takes
1.09\,s, refilling weights takes 0.85\,s, restoring NIXL connectivity takes
2.50\,s, and \texttt{resize} takes 2.55\,s. The remaining control steps are
small in both directions. Thus, expert re-layout dominates \sys scale-down,
while scale-up also pays for restoring workers, weights, and connectivity.

\begin{figure}[t]
  \centering
  \includegraphics[width=\columnwidth]{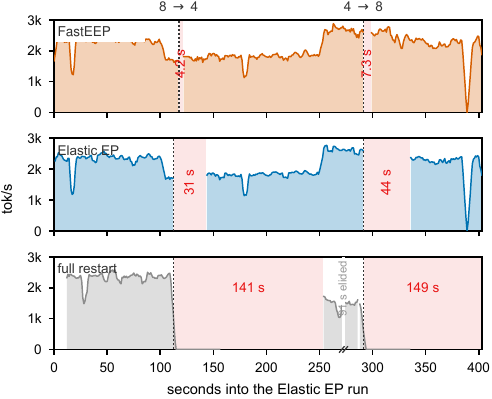}
  \caption{Throughput during an $8\rightarrow4\rightarrow8$ scaling cycle
  under load (5\,s rolling mean). Red bands mark zero-output intervals, whose
  durations measure service interruption.}
  \label{fig:trace}
\end{figure}

\subsection{Scaling While Serving Requests}
\label{sec:eval:load}
\label{sec:eval:roundtrip}

We send ShareGPT requests with a closed-loop concurrency of 128 for the first
100\,s, 64 from 100 to 250\,s, and 128 thereafter. The service follows
$8\rightarrow4\rightarrow8$. Figure~\ref{fig:trace} shows the throughput
trace; all transitions are issued through the public HTTP API.
\textbf{\sys substantially shortens serving interruptions.}
For the $8\rightarrow4$ and $4\rightarrow8$ transitions, the
zero-output windows are 4.2\,s and 7.3\,s for \sys, compared with 31\,s and
44\,s for Elastic EP and 141\,s and 149\,s for restart. Reusing worker,
communication, and CUDA Graph state therefore allows serving to resume much
sooner than rebuilding the runtime.

\textbf{Throughput tracks the active EP size.}
After scale-down, throughput settles at a lower level because fewer workers
remain active; after scale-up, it returns to the higher range. Short dips
outside the marked intervals reflect workload and runtime variation rather
than additional scaling outages.

\subsection{Scaling Multiple Models Simultaneously in MaaS}
\label{sec:eval:multi}
\label{sec:eval:mixed}
\label{sec:eval:power}

We colocate Qwen3-30B-A3B and DeepSeek-V2-Lite under independent demand.
In Figure~\ref{fig:multi}, Service~1 scales from EP 8 to 4 and releases four
GPUs to Service~2. When its demand returns, Service~2 first scales from four
GPUs to two, allowing Service~1 to grow from EP 4 to 6. Neither instance is
reconstructed. Light-pink intervals are asynchronous drains that overlap
serving; dark-pink intervals have no token output. The first $8\!\rightarrow\!4$
transition drains for 3\,s and reconfigures in 4.4\,s, while driver-reported
free memory rises from 34\,191 to 64\,613\,MiB.

\textbf{Cross-service scaling imposes an ordering dependency.}
Released pages return to the node allocator rather than transferring directly
between processes, so dependent transitions must be ordered. Service~1's
$4\!\rightarrow\!6$ scale-up (6.9\,s) therefore follows Service~2's
$4\!\rightarrow\!2$ scale-down (4.2\,s).

\begin{figure}[t]
  \centering
  \includegraphics[width=\columnwidth]{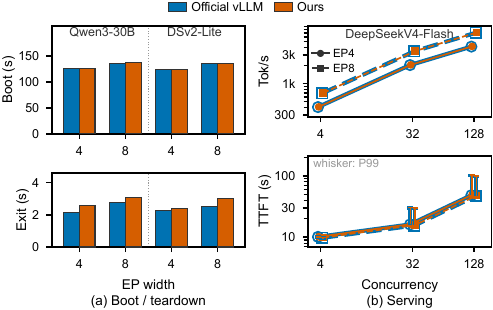}
  \caption{Boot, teardown, and serving performance of \sys and unmodified
  vLLM at EP 4 and EP 8.}
  \label{fig:vllmsdiff}
\end{figure}

\subsection{Performance Impact of \sys}
Figure~\ref{fig:vllmsdiff} shows that \sys preserves unmodified vLLM's
steady-state throughput at EP 4 and EP 8. Inactive-worker buffers move to
CPU memory, while dormant workers run synchronization-only dummy iterations
and launch no model computation. Partial reuse therefore adds no work to the
steady-state inference path. Fully asynchronous scaling can transiently
contend with inference while it prefetches or offloads buffers and experts,
but moving these transfers before the brief global pause reduces scaling
latency.

%% file: 07-discussion.tex

\section{Discussion}
\label{sec:discussion}

\paragraph{The space--time trade-off.}
Fast scaling trades memory for reconstruction time. Rather than retaining a
complete instance or preloaded replica, \sys manages workers, communicators,
and memory regions independently. For state that is expensive to rebuild, it
retains only the metadata needed for reuse, while releasing or offloading data
resources unused by the active configuration. This enables fast transitions
without the memory cost of an entire inactive configuration.

\paragraph{Prioritizing GPU resource reclamation.}
Persistence does not imply GPU residency. \sys maintains a communication group
at the maximum world size, but after each resize it offloads inactive GPU data
buffers to CPU memory (2--4\,GB, excluding the KV cache) and retains only the metadata needed for reuse.
Initializing the system at the maximum world size therefore does not require
the corresponding GPU resources to remain occupied throughout execution.

\paragraph{Scope and scalability.}
Unlike systems that recover failed ranks or switch entire
models~\cite{sglang-rankfail,aegaeon}, \sys focuses on changing
the EP width of a running MoE instance.
We evaluate \sys on a two-node, 16-GPU cluster with NCCL communication and
inter-node RDMA. Although our evaluation is limited to 16 GPUs, the design
scales to larger clusters by replacing centralized connection reconstruction
with distributed state offloading. This decentralized approach should scale
more favorably than existing scaling mechanisms. Our implementation is open
source and has been submitted to vLLM as a PR.
 
\paragraph{Limitations.}
\sys scales only within the maximum topology initialized at startup. Scale-up
requires sufficient available GPU memory, and larger topology changes still
require reconstruction.

%% file: 08-related.tex

\section{Related Work}
\label{sec:related}

\paragraph{Elastic MoE serving.}
vLLM Elastic EP~\cite{vllm-eep-blog} changes world size but rebuilds
communicators. ElasticMoE~\cite{elasticmoe}, evaluated on Huawei's Ascend NPUs,
offloads only model weights and KV caches yet remains fundamentally
instance-level: scale-up switches to a preinitialized instance and requires
warm-up.
MoEless~\cite{moeless} scales individual experts as serverless functions.
Rank-failure recovery~\cite{sglang-rankfail} preserves CUDA Graphs through
peer masking and expert remapping, without reclaiming GPU memory.
\sys supports bidirectional rank scaling within one preinitialized maximum
topology, reusing graphs and communication state while reclaiming inactive
ranks' physical memory.

\paragraph{Instance scaling and scheduling.}
Instance-level systems optimize loading and cold
starts~\cite{blitzscale,lambdascale,serverlessllm,hydraserve,warmserve} or
multiplex models and requests~\cite{alpaserve,muxserve,aegaeon,llumnix,melange}.
EP size offers an additional allocation knob for goodput-oriented
schedulers~\cite{pollux}; elastic sequence parallelism and memory pools
provide complementary dimensions~\cite{loongserve,memserve}.
Autoscaling policies decide when to resize~\cite{chiron,enova,tokenscale,
taming-chaos,dynamollm,hfx}; our mechanism supplies the transition and supports
either side of prefill/decode disaggregation~\cite{distserve,splitwise,
mooncake}, although we evaluate a non-disaggregated deployment.

\paragraph{GPU virtual memory.}
vAttention and KunServe grow KV capacity with VMM~\cite{vattention,kunserve};
Prism, WarmServe, and vLLM sleep mode release pages without changing virtual
addresses~\cite{prism,warmserve}. Related abstractions manage
tensors, defragmentation, memory ballooning, and MoE experts~\cite{vtensor,
gmlake,ellm,expertweave}.
Foundry~\cite{foundry} uses deterministic addresses for graph transfer across
processes; FlashBoot~\cite{flashboot} accelerates weight loading and avoids
NCCL group formation.
\sys combines page release with graph replay and RDMA registration renewal
across repeated EP resizes.

%% file: 09-conclusion.tex

\section{Conclusion}
\label{sec:conclusion}

We presented \sys, a system for fast intra-instance EP scaling in MoE
serving. Through partial state reuse, \sys preserves workers, stable virtual
addresses, and the startup communication topology while selectively reclaiming
and reconfiguring GPU resources. 
This design substantially reduces scaling
latency and allows a running instance to adjust its GPU allocation without
reconstruction. 
\sys enables MoE services to adapt to changing
demand beyond fixed instance boundaries.